\documentclass[aps,twocolumn,pra,superscriptaddress,floatfix,showpacs]{revtex4-1}
\usepackage{amsmath}
\usepackage{graphicx}
\usepackage{subfigure}
\usepackage[colorlinks=true,linktoc=page,linkcolor=blue,citecolor=blue,urlcolor=blue]{hyperref}
\usepackage{color}
\usepackage{epstopdf}
\usepackage{float}

\def\lsim{\mathrel{\mathpalette\gl@align<}}
\def\gsim{\mathrel{\mathpalette\gl@align>}}
\def\gl@align#1#2{\lower.6ex\vbox
{\baselineskip\z@skip\lineskip\z@
\ialign{$\m@th#1\hfil##\hfil$\crcr#2\crcr\sim\crcr}}}

\makeatother

\newcommand\ba{\begin{eqnarray}}
\newcommand\ea{\end{eqnarray}}
\newcommand\be{\begin{equation}}
\newcommand\ee{\end{equation}}

\usepackage[normalem]{ulem}
\usepackage{sidecap,tikz}
\DeclareRobustCommand{\orcidicon}{\hspace{-1.0mm}
	\begin{tikzpicture}
	\draw[lime, fill=lime] (0.0,0.0) 
	circle [radius=0.15] 
	node[white] {{\fontfamily{qag}\selectfont \tiny \,ID}};
	\draw[white, fill=white] (-0.0525,0.095) 
	circle [radius=0.007];
	\end{tikzpicture}
	\hspace{-3.0mm}
}
\foreach \x in {A, ..., Z}{\expandafter\xdef\csname orcid\x\endcsname{\noexpand\href{https://orcid.org/\csname orcidauthor\x\endcsname}
		{\noexpand\orcidicon}}
}

\begin{document}

\title{Persistent anomaly in dynamical quantum phase transition in long-range non-Hermitian $p$-wave Kitaev chain}

\author{Debashish Mondal\orcidC{}}
\email{debashish.m@iopb.res.in}
\affiliation{Institute of Physics, Sachivalaya Marg, Bhubaneswar-751005, India}
\affiliation{Homi Bhabha National Institute, Training School Complex, Anushakti Nagar, Mumbai 400094, India}

\author{Tanay Nag\orcidB{}}
\email{tanay.nag@hyderabad.bits-pilani.ac.in}
\affiliation{Department of Physics, BITS Pilani-Hydrabad Campus, Telangana 500078, India}


\begin{abstract}
Considering a non-Hermitian version of $p$-wave Kitaev chain in the presence of additional second nearest neighbour tunneling, we study  dynamical quantum phase transition~(DQPT) which accounts for the vanishing Loschmidt amplitude.
The locus of the Fisher's zero traces a continuous path on the complex time plane for the Hermitian case while it becomes discontinuous for non-Hermitian cases. This further leads to the half-unit jumps in the winding number characterizing a dynamical topological aspect of DQPT for non-Hermitian Hamiltonian. Uncovering the interplay between non-Hermiticity and long-range tunneling, we find these features to be universally present irrespective of the additional second nearest neighbour tunneling terms as long as non-Hermiticity is preserved. 
\end{abstract}

\maketitle

{\bf Keywords:} Dynamical quantum phase transitions, non-Hermitian effects, $p$-wave Kitaev chain, non-equilibrium phenomena


\section{Introduction}

The phenomena of phase transition 
in the thermodynamic limit can be understood through the non-analyticities in the free-energy density, equivalently labeled by the zeros of the partition function i.e.,  Fisher zeros in complex 
parameter planes \cite{fisher1967theory,LYF1, LYF2}. This is how   magnetic and liquid-gas phase  transitions  in 
magnetic field-temperature and pressure-temperature planes, respectively, can be apprehended.
The dynamical quantum phase transition~(DQPT), on the other hand, is associated with the singular nature of the dynamical free-energy density at certain critical times. Interestingly, DQPT can be interpreted by vanishing nature of Loschmidt amplitude in complex time plane such that time-evolved state becomes orthogonal to the initial state \cite{heyl13,PhysRevB.87.195104,PhysRevB.90.125106,PhysRevLett.113.265702,PhysRevLett.115.140602,Heyl_2018,Bhattacharya17,Jafari19a,Uhrich20,SMB_PRL,cao2023aperiodic}. The DQPTs are further identified by the jumps in winding number, acting like a  dynamical topological marker, that characterizes the topological properties of the real-time dynamics.
Apparently, there has not been any concrete connection established so far between the quantum critical point (QCP) and the emergence of DQPT  \cite{heyl13,Budich16, SS,PhysRevB.92.104306,Divakaran16,Dutta17,Vajna14,Schmitt15,Halimeh17,Silva18,Halimeh20c,Hashizume22,Lang18,Homrighausen17,rossi2022non,mishra2020disordered} rendering the applicability of DQPT in various theoretical scenarios \cite{PhysRevB.89.161105, PhysRevB.92.235433,PhysRevB.89.125120,Halimeh17,Modak21,Abdi19,PhysRevB.103.064306,PhysRevResearch.4.013002,Zamani20,Jafari22,Jafari21a,Jafari21,Jafari17,zhou2021floquet,Yang19,Kosior18,Kosior18b,Halimeh_sugg_2019,Halimeh_sugg_2017} and experimental setups  \cite{PhysRevLett.119.080501,Nie20,flaschner2018observation}.

The Hermitian version of DQPT is very recently generalized to non-Hermitian framework \textcolor{black}{where the non-Hermiticity is mimicked by the  imaginary self-energy namely, life-time of the underlying quasi-particles. Therefore, the above framework} provides an equivalent description of 
open quantum system \cite{Bergholtz19,Yang21}, quasiparticles system with finite lifetime \cite{kozii2017non,Yoshida18,Shen18,Gou20,Zeuner15,weimann2017topologically,Weiwei18,Gao20}. Interestingly, non-Hermitian systems host, instead of QCPs, exceptional points
where eigenvectors, associated with the degenerate bands
fuse~\cite{Bergholtz21,ghatak2019new,ashida2020non,Kawabata19,Shen18}. This along with the non-unitarity in the time evolution non-trivially modify the time evolution of the winding number  in non-Hermitian case \cite{Zhou1,Zhou_2021,PhysRevA.105.022220,Hamazaki2021,Mondal_2}. To be precise, one comes across half-unit jumps in the winding number, unlike the Hermitian counterpart where unit jumps are  observed only ~\cite{Mondal_1,jing2023biorthogonal}. The above studies are limited to the first nearest neighbour models. Therefore, motivated by the above findings, we here focus on the aspect of whether the half-unit jumps continue to persist for a non-Hermitian model with second nearest neighbour connections. This allows us to explore the
interplay between non-Hermiticity and long-range character of the model.

\begin{figure*}[] 
\includegraphics[trim=0.0cm 0.0cm 0.0cm 0.0cm, clip=true, height=!,width=2.0\columnwidth]{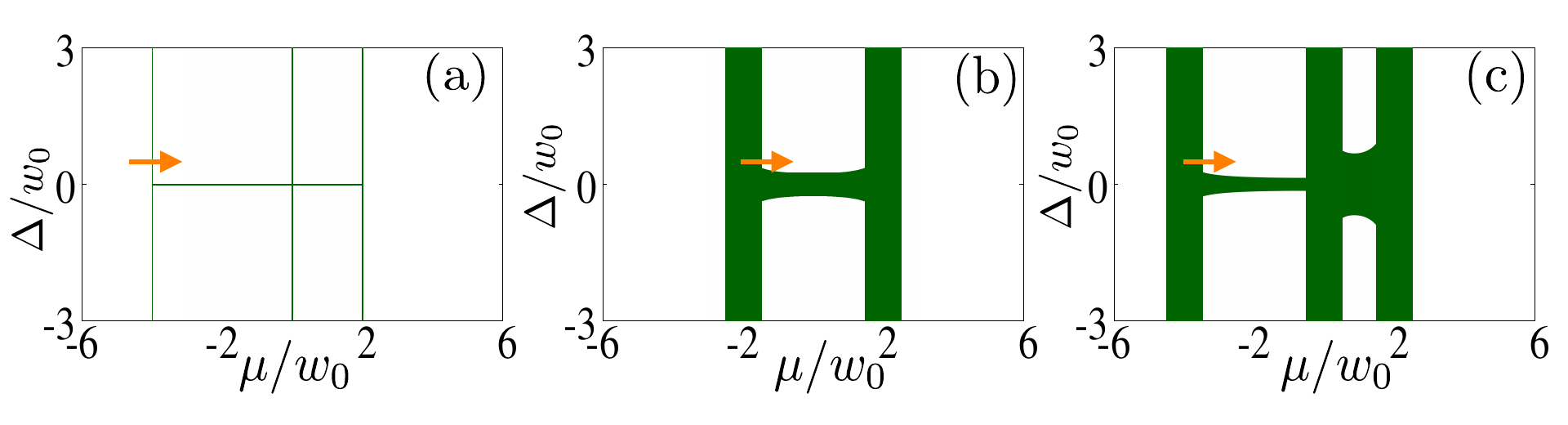}
\caption{The schematic phase diagram of the model Hamiltonian in Eq.~(\ref{eq:Momentum_Ham}) is shown in panel (a,b,c) for ($\gamma=0$, $\beta =1$), ($\gamma=1$, $\beta=0$) and ($\gamma=1$, $\beta =1$), respectively.  The  white regions represent  the gapped  phases  while green regions are gapless. Orange arrows represent paths for sudden quench. For (b,c) panels quench paths cross one exceptional line while  one critical line for (a). We consider $w_{0}=1$. } \label{fig:phase}
\end{figure*}

We consider an extended $p$-wave superconductor chain with second nearest neighbour tunneling where the range \textcolor{black}{of} the hopping as well as superconductivity both increase. \textcolor{black}{To be precise, the second nearest neighbor hopping and second nearest neighbor superconducting pairing  are included in addition to their first nearest neighbor counterparts in the $p$-wave superconductor chain.} The Hermiticity breaking term is included in the superconducting part which induces a finite gapless region in the phase diagram while long-range nature brings new gapless line, see Fig, \ref{fig:phase}. We first study the simplest  Hermitian case by including the second nearest neighbour tunneling where 
unit jumps, associated with continuous profile of Fisher's zeros, in winding numbers are observed, see Fig.~\ref{fig:results}(a). Next, we add the non-Hermitian term while excluding the second nearest neighbour tunneling and  we find discontinuous profile of Fisher's zeros yielding half-unit jumps, see Fig.~\ref{fig:results}(b). We finally include the second nearest neighbour tunneling along with non-Hermiticity to show that the half-unit jump is indeed a non-Hermitian effect while long-rangeness changes the location of Fisher's zeros, see Fig.~\ref{fig:results}(c).

The paper is organized as follows. We discuss the details of the models and the corresponding phase diagrams in Sec. \ref{model}. Next, we demonstrate the theory behind the non-Hermitian DQPT in Sec.  \ref{theory}. We illustrate our main findings on the profile of  Fisher's zero and dynamical winding number in Sec. \ref{results}. At the end, we conclude in Sec. \ref{conclusions}.

\section{Model }
\label{model}

We start with the non-Hermitian analog of a 1D $p$-wave superconductor  with first and second nearest neighbour hopping as well as superconducting pairing \cite{Mondal_1,Mondal_2}. Considering the particle-hole basis $\psi_k= (c_k, c^{\dagger}_{-k})^T$,  the model Hamiltonian is block-diagonal in terms of the momentum modes as given by  \textcolor{black}{$H(\gamma)= \sum_{k} \frac{1}{2}\psi^{\dagger}_k { H}_{k}(\gamma) \psi_k $} where two-level ${ H}_{k}(\gamma)$ takes the form 
\cite{DeGo1,Manisha1,Rajak1} 
\begin{eqnarray}
{ H}_{k}(\gamma) =&& \Big[2 \Delta (\sin k+\beta \sin 2k) +\frac{i\gamma}{2}\Big]\sigma_{y} - \nonumber \\
&&\Big[2 w_0  (\cos k+ \beta \cos 2k) +\mu  
\Big]\sigma_{z}  =\vec{h}_k \cdot \vec{\sigma}
\label{eq:Momentum_Ham},
\end{eqnarray}
where $w_0$ ($\beta w_0$), and  $\Delta$  ($\beta \Delta$) represent the first (second) nearest neighbour hopping amplitude and superconducting gap, respectively.   $\mu$ denotes the chemical potential. For $\beta=0$ and $\gamma=0$, the model corresponds to the Hermitian $p$-wave Kitaev chain  with first nearest neighbour tunneling \cite{Kitaev_2001}.  Note that $\vec{h}_k=\{h^y_k,h^z_k\}$ and $\vec{\sigma}=\{\sigma_y,\sigma_z\}$.  Here, the non-Hermitian term $i\gamma/2$, added to the superconducting gap, can be thought of as lossy superconductivity which may arise due to spatially separated pairing processes~\cite{Shi_parining2023}. \textcolor{black}{To be precise, the non-Hermitian part of the superconductivity corresponds to an onsite $s$-wave like pairing with exactly opposite pairing strengths between the Hermitian partners.  
The underlying real space representation, consisting of $N$ lattice sites in a one-dimensional chain,  of the  model Hamiltonian (\ref{eq:Momentum_Ham}) is given by \begin{eqnarray}
  &H(\gamma)&= -w_{0} \sum_{j=1}^{N-1}\left( c_{j}^{\dagger} c_{j+1}+  c_{j+1}^{\dagger} c_{j} \right) -\mu  \sum_{j=1}^{N} \left(c_{j}^{\dagger} c_{j} -\frac{1}{2}\right) \nonumber\\
  &+&\Delta \sum_{j=1}^{N-1}  \left(c_{j+1}^{\dagger}c_{j}^{\dagger} +c_{j}c_{j+1} \right) + \frac{\gamma}{4} \sum_{j=1}^{N} 
 \left(c_{j}c_{j} -c_{j}^{\dagger}c_{j}^{\dagger} \right) \nonumber\\
  &-& w_{0}\beta \sum_{j=1}^{N-2} \left( c_{j}^{\dagger} c_{j+2}+  c_{j+2}^{\dagger} c_{j} \right)  \nonumber \\
  &+& \Delta \beta \sum_{j=1}^{N-2}  \left(c_{j+2}^{\dagger}c_{j}^{\dagger} +c_{j}c_{j+2} \right).
\label{eq:real_ham}
\end{eqnarray} 
Here, $c_j (c_j^{\dagger})$ denotes the fermionic annihilation (creation) operator at $j$-th site.}

As discussed earlier, the non-Hermiticity can be thought of as an effect coming from the leads connected to an open $p$-wave Kitaev chain. For simplicity, we consider either $\beta=1$ or $\beta=0$ for the rest of our analysis. 
The Hamiltonian~(\ref{eq:Momentum_Ham}) becomes gapless for a particular momentum $k_{*}$, if the real parts of its energy eigenvalues become zero i.e.
\begin{equation}
  \left[ 2w_{0} (\cos k_{*} +\cos 2k_{*})+\mu\right]^{2} +  4 \Delta^{2} (\sin k_{*} +\sin2k_{*})^{2} 
    -\frac{\gamma^{2}}{4}=0. \label{eq:gapless}  
\end{equation}
The corresponding phase diagram~(schematic) is shown in Fig.~\ref{fig:phase} for three different cases.
For Hermitian counterpart with $\gamma=0$ and $\beta=1$, the gapless lines, consisting of QCPs, are given by $\mu=-4w_{0}$, $\mu=2w_{0}$, $\mu=0$ and $\Delta=0$, see Fig.~\ref{fig:phase}(a). These lines separate different gapped phases with their distinct topological properties \cite{DeGo1}. Upon adding the non-Hermitian factor $\gamma \ne 0$ along with $\beta=1$, the above gapless lines expand into extended gapless regions, depicted by green color in Fig. ~\ref{fig:phase}(c) within which the real part of the energy vanishes. The gapless region is bounded by the exceptional points leading to 
vertical exceptional lines $\mu=-4w_{0}\pm \gamma/2$, $\mu=2w_{0}\pm \gamma/2$, and $\mu=\pm \gamma/2$ and the horizontal exceptional lines $\Delta=\pm \frac{\gamma}{4 (2 + 2 \alpha_{\pm} -\mu^2/4 w_0^2)^{1/2}}$ with $4 \alpha_{\pm}=-1 \pm \sqrt{9-4\mu/w_0}$ such that $\alpha_{\pm}$ remains real and $-1<\alpha_{\pm}<1$. 
The horizontal (vertical) exceptional lines are obtained when  $\cos k_* + \cos 2 k_* = -\frac{\mu}{2 w_{0}}$ ($\sin k_{*} +\sin2k_{*}=0$). In the absence of second nearest neighbour connections for the non-Hermitian model i.e., $\beta=0$ and $\gamma \ne 0$, phase boundaries  are given by $\mu=\pm 2 w_{0} \pm \gamma/2$ and $\Delta=\pm \gamma/ \sqrt{16-\frac{4\mu^{2}}{w_{0}^{2}}}$, see Fig.~\ref{fig:phase}(b)). Therefore, the non-Hermiticity causes the transformation of critical lines to the critical regions while second neighbour tunneling is solely responsible for generating new critical lines on the phase diagram.

\section{Theory}\label{theory}
The Loschmidt amplitude, associated with a given momentum mode, reads as the  overlap between the initial state and  the time-evolved  state \textcolor{black}{$g_k(t)=\langle \psi_{k,i} | \exp(-i H_{k,f} t  )| \psi_{k,i} \rangle $}~\cite{Zhou1,Mondal_1}. Here, $|\psi_{k,i}\rangle$ denotes the 
initial state i.e., an eigenstate of 
\textcolor{black}{initial} Hamiltonian \textcolor{black}{$H_{k,i}=\vec{h}_{k,i}.\vec{\sigma}=h_{k,i} \hat{h}_{k,i}.\vec{\sigma}$} which is time-evolved by a final Hamiltonian \textcolor{black}{ $H_{k,f}=\vec{h}_{k,f}.\vec{\sigma}=h_{k,f} \hat{h}_{k,f}.\vec{\sigma}$} following a sudden quench. Since we deal with non-Hermitian Hamiltonians, we always work on the bi-orthogonalized basis such that $\sum_l|\psi^l_{k,i(f)}\rangle \langle\psi^l_{k,i(f)}|=I$ and 
$\langle\psi^l_{k,i(f)}| \psi^m_{k,i(f)}\rangle=\delta_{ml}$. 
The Loschmidt amplitude is thus given by 
\begin{equation}
 g_k(t)=\cos(h_{k,f}t)-i\sin(h_{k,f}t)\langle\psi_{k,i}|\frac{{H}_{k,f}}{h_{k,f}}|\psi_{k,i}\rangle. \label{eq:g_k}
\end{equation}
As discussed earlier, the DQPTs take place when Loschmidt amplitude vanishes in complex time plane, resulting in Fisher's zeros as given by~\cite{Zhou1,Mondal_1}
\begin{equation}
z_{n,k}=i\frac{\pi}{h_{k,f}}\left(n+\frac{1}{2}\right)+\frac{1}{h_{k,f}}{\rm arctanh}\langle\psi_{k,i}|\frac{{ H}_{k,f}}{h_{k,f}}|\psi_{k,i}\rangle,\label{eq:LYF}
\end{equation}
where $z_{n,k}=it$ and $n \in Z$. The dynamical order parameter namely,  winding number is given by~\cite{Budich16}
\begin{equation}
\nu(t)=\frac{1}{2\pi}\oint_{BZ}dk\left[\partial_{k}\phi_{k}^{{\rm G}}(t)\right].\label{eq:WN}
\end{equation}
Here, the geometric phase  is given by $\phi_k^G(t)={\phi}_{k}^{\rm tot}(t)-\phi_k^{\rm dyn}(t)$, with total phase  ${\phi}_{k}^{\rm tot}(t)= - i \ln \left[\frac{g_k(t)}{|g_k(t)|}\right]$ and the dynamical phase is found to be ~\cite{Zhou1,Gong1}
\begin{eqnarray}
\phi_{k}^{{\rm dyn}}(t)=  &-&\int_{0}^{t}dt^{\prime}\frac{\langle\psi_{k,i}(t^{\prime})|{ H}_{k,f}|\psi_{k,i}(t^{\prime})\rangle}{\langle\psi_{k,i}(t^{\prime})|\psi_{k,i}(t^{\prime})\rangle} \nonumber\\
&+&\frac{i}{2}\ln\left[\frac{\langle\psi_{k,i}(t)|\psi_{k,i}(t)\rangle}{\langle\psi_{k,i}(0)|\psi_{k,i}(0)\rangle}\right],\label{eq:dyn} 
\end{eqnarray}
with $|\psi_{k,i}(t)\rangle=e^{-i{H}_{k,f} t }|\psi_{k,i}\rangle$ and $\langle\psi_{k,i}(t)|=\langle\psi_{k,i}| e^{i {  H^\dagger}_{k,f} t}$ are the time-evolved  right and left eigenstates of ${H}_{k,i}$ respectively.

\begin{figure*}[t] 
\includegraphics[trim=0.0cm 0.0cm 0.0cm 0.0cm, clip=true, height=!,width=2.0\columnwidth]{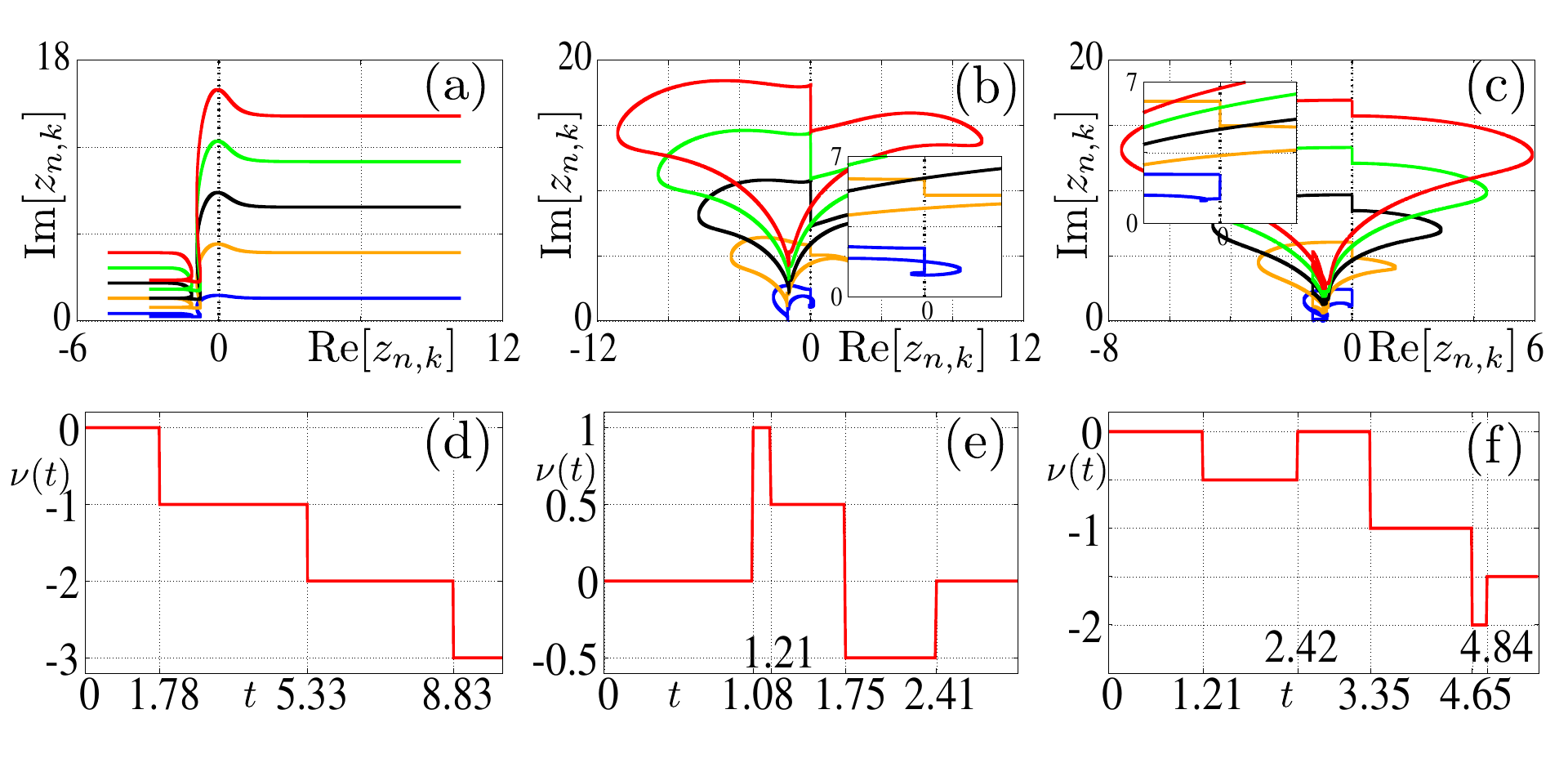}
\caption{Lines of Fisher's zeros $z_{n,k}$ with $n=0$(blue), $1$ (orange), $2$ (black), $3$ (green), $n=4$(red) computed from Eq.~(\ref{eq:LYF}) are depicted here in panels (a,b,c) corresponding to quench path shown in the Fig~\ref{fig:phase}(a,b,c) with ($\gamma=0$, $\beta =1$), ($\gamma=1$, $\beta=0$) and ($\gamma=1$, $\beta =1$), respectively.  The  dynamical winding number $\nu(t)$, obtained from  Eq. (\ref{eq:WN}), for (a,b,c) as a function of time $t$ are shown in (d,e,f), respectively. The  unit~(half-unit) change in winding number is connected to the continuous (discontinuous) lines of Fisher's zeros over the   \textcolor{black}{complex} time plane as shown in the upper panel. Parameters ($\mu_{i},\mu_{f},\Delta$) corresponding to (a,d),  (b,e), and (c,f) are as follows (-4.5,-3,0.4), (-2,-1,0.4), and (-4,-3,0.4), respectively.} \label{fig:results}
\end{figure*}


\section{results}
\label{results}
We now demonstrate the DQPT for three different cases namely, Hermitian second nearest neighbour, non-Hermitian first nearest neighbour, and non-Hermitian second nearest neighbour model, aiming to understand the interplay between the non-Hermiticity and long-range tunneling. We below demonstrate a particular representative situation from each of the models to highlight our findings. 
Note that there exist several other quench paths and our findings are qualitatively unaltered.



We begin our analysis by considering the second nearest neighbour hopping and superconductivity in the Hermitian model (Eq.~(\ref{eq:Momentum_Ham}) with $\beta=1$ and $\gamma=0$) to explore long-range effect on DQPT.  Following the 
sudden quench from one to the other gapped phase through a critical line at $\mu=-4w_{0}$, shown in  Fig.~\ref{fig:phase}(a) by orange arrow, 
we find the lines of Fisher's zeros continuously cross the imaginary axis. This is depicted in Fig.~\ref{fig:results}(a) while the integer jumps in the winding number at time $t \approx 1.78, 5.33,..$, shown in Fig.~\ref{fig:results}(d), are associated with different values of $n$.  These critical times are associated with the vanishing nature of the real part in $z_{n,k}$ as discussed in Eq. (\ref{eq:LYF}).
Since the lines of Fisher's zero cross the imaginary axis once, we only find monotonic change in the winding number.  
Therefore, the DQPT profile does not qualitatively change upon adding the   second neighbour tunneling   compared to the first nearest neighbour tunneling \cite{Mondal_1}. 



We now extend our analysis to the non-Hermitian case while the first nearest neighbour tunneling in the Hamiltonian (Eq.~(\ref{eq:Momentum_Ham}) with $\beta=0$ and $\gamma=1$) is taken into consideration only. This enables us to  study the exclusive effect of non-Hermiticity on the DQPT. We adopt a  quench through an exceptional line $\mu=-2w_{0}+\gamma/2$ when the  Hamiltonian is suddenly changed from a non-Hermitian gapless phase to a gapped phase as shown by orange arrow in Fig.~\ref{fig:phase}(b).  We find  that the lines of Fisher's zeros exhibit both continuous as well as discontinuous crossings over the imaginary axis, see Fig.~\ref{fig:results} (b). Interestingly, the line of Fisher's zeros crosses the imaginary axis twice out of which the discontinuous profile is noticed once. As a result, the winding number displays non-monotonic jumps associated with multiple crossings while continuous and discontinuous crossings cause unit and half-unit jumps, respectively, see Fig.~\ref{fig:results} (e). Remarkably, the half-unit jump is a unique feature of non-Hermitian Hamiltonian and is completely absent in the Hermitian counterpart \cite{Mondal_1}.

The discontinuous profile of Fisher's zeros at 
${\rm Re}[z_{n,k}]=0$ can be understood by a complex nature of the quantity $n$ in Eq. \ref{eq:LYF}. It has been shown that in non-Hermitian topological phases, the non-Bloch complex form of momentum is essential to understand the emergence of boundary modes \cite{PhysRevB.106.L140303}. Likewise, we can also naively think that the complex momentum induces a complex $n$ leading to discontinuous profiles for the lines of  Fisher's zeros. However, the exact determination of $n$ from an accurate non-Bloch momentum is an open question. By contrast, the Bloch nature of the real momentum results in real-valued $n$. Therefore, non-Hermitian DQPT is characteristically different from its  Hermitian analog.



At the end, we contemplate a situation where non-Hermiticity and second nearest tunneling are both present simultaneously in the Hamiltonian (Eq.~(\ref{eq:Momentum_Ham}) with $\beta=1$ and $\gamma=1$) such that their mutual interplay can be captured via the occurrence of DQPT. We follow the quench path passing through the exception line $\mu=-4w_{0}+\gamma/2$ from the initial non-Hermitian gapless phase to the gapped phase, as shown by orange arrow in Fig. ~\ref{fig:phase}(c). The lines of the Fisher's zero show combinations of discontinuous and continuous crossings of the imaginary axis, see Fig.~\ref{fig:results}(c). This is 
qualitatively similar to what is already noticed in Fig.~\ref{fig:results}(b). The long-range tunneling can only change the shape of the lines of Fisher's zero without altering their distinct discontinuous features. Therefore, the non-monotonic nature as well as half-unit jumps are visible in the dynamical winding number, see Fig.~\ref{fig:results}(f).  A close inspection suggests that the continuous jumps appear at $t\approx 3.35, 4.65,..$ and while discontinuous jumps take place at $t \approx 1.21, 2.42,4.84,...$. Therefore, the second nearest neighbour tunneling does not destroy the non-Hermiticity mediated half-unit jumps in the winding number.


 \section{conclusions}\label{conclusions}
We consider non-Hermitian analog of one dimensional $p$-wave Kitaev chain with additional second nearest neighbour hopping and superconducting coupling. The second nearest neighbour introduces new critical lines while non-Hermiticity brings in extended gapless regions around the critical lines.  We find that the characteristics of DQPT markedly change due to non-Hermiticity where the discontinuous profiles of Fisher's zeros and half-quantized jumps in the winding number are observed. These features are absent for the Hermitian counterpart irrespective of the second nearest neighbour tunneling and hence can be inferred as the unique signatures of non-Hermitian DQPT \cite{Mondal_1}.  In the future, on a similar line one can study the effect of power-law hopping and superconducting coupling on the non-Hermitian DQPT.


\section*{acknowledgement}
We would like to dedicate this work to Prof. Amit Dutta whose untimely demise is a great loss for the community. Being his Ph.D. student, I (TN) always wanted to work with him on non-Hermitian DQPT once I returned to India. Unfortunately, this did not take place due to the unfortunate event.  We are thankful to Heiko Rieger and Eduardo Hernandez, former and
present Editor-in-Chief of European Physical Journal B (EPJB), for taking the initiative of
this Topical Issue on “Quantum phase transitions and open quantum systems: A tribute to
Prof. Amit Dutta”, in memory of their one long time Editor. We would like to thank the
Guest Editors of this Special Issue of EPJB, Uma Divakaran, Ferenc Igloi, Victor Mukherjee
and Krishnendu Sengupta for kind invitation to contribute in it. DM acknowledges SAMKHYA (High-Performance Computing Facility provided by the Institute of Physics,
Bhubaneswar) for numerical computations. We thank to Arjit Saha for useful discussions. TN  acknowledges the NFSG “NFSG/HYD/2023/H0911” from BITS Pilani.

{\bf Data availability statement:} The data may be available on request to the corresponding
author.

{\bf Conflict of interest statement:} We declare that this manuscript is free from any conflict
of interest. The authors have no financial or proprietary interests in any material discussed in
this article.

{\bf Funding statement:} No funding was received particularly to support this work.

{\bf Authors’ contributions:} Tanay Nag conceived the idea, analyzed the results and wrote the manuscript. Debashish Mondal did all the numerical calculations, prepared the figures, analysed the results and partially wrote the manuscript.

\bibliography{bibfile}{}

\end{document}